\documentclass[useAMS,usenatbib]{mn2e}
\title[CLASS B2108+213: A new wide separation gravitational lens system]{CLASS B2108+213: A new wide separation gravitational lens system}
\author[J. P. McKean et al.]
{J. P. McKean,$^{1,2}$\thanks{Email: mckean@physics.ucdavis.edu} I. W. A. Browne,$^1$ N. J. Jackson,$^1$ L. V. E. Koopmans,$^3$  
\newauthor M. A. Norbury,$^1$ T. Treu,$^4$\thanks{Hubble Fellow} T. D. York,$^1$ A. D. Biggs,$^5$ R. D. Blandford,$^6$ 
\newauthor A. G. de Bruyn,$^{3,7}$ C. D. Fassnacht,$^2$  S. Mao,$^1$ S. T. Myers,$^8$  T. J. Pearson,$^{9}$  
\newauthor P. M. Phillips,$^1$ A. C. S. Readhead,$^{9}$ D. Rusin$^{10}$ and P. N. Wilkinson$^1$\\
$^1$University of Manchester, Jodrell Bank Observatory, Macclesfield, Cheshire SK11 9DL\\
$^2$Department of Physics, University of California, Davis, CA 95616, USA\\
$^3$Kapteyn Astronomical Institute, Postbus 800, NL-9700 AV Groningen, the Netherlands\\
$^4$Department of Astronomy and Astrophysics, University of California, Los Angeles, CA 90095, USA\\
$^5$Joint Institute for VLBI in Europe, Postbus 2, 7990 AA Dwingeloo, the Netherlands\\
$^6$KIPAC, Stanford University, 2575 Sand Hill Road, Menlo Park, CA 94025, USA\\
$^7$Netherlands Foundation for Research in Astronomy, Postbus 2, 7990 AA Dwingeloo, the Netherlands\\
$^8$National Radio Astronomy Observatory, P.O. Box 0, Socorro, NM 87801, USA\\
$^{9}$California Institute of Technology, Pasadena, CA 91125, USA\\
$^{10}$Department of Physics and Astronomy, University of Pennsylvania, 209 South 33rd Street, Philadelphia, PA 19104, USA}

\begin{document}

\date{Accepted 2004 August 31. Received 2004 August 31; in original form 2004 August 31}

\pagerange{\pageref{firstpage}--\pageref{lastpage}} \pubyear{2004}

\maketitle

\label{firstpage}

\begin{abstract}
We present observations of CLASS B2108+213, the widest separation gravitational lens system discovered by the Cosmic Lens All-Sky Survey. Radio imaging using the VLA at 8.46~GHz and MERLIN at 5~GHz shows two compact components separated by 4.56~arcsec with a faint third component in between which we believe is emission from a lensing galaxy. 5-GHz VLBA observations reveal milliarcsecond-scale
structure in the two lensed images that is consistent with gravitational lensing. Optical emission from the two lensed images and two lensing galaxies within the Einstein radius is detected in {\it Hubble Space Telescope} imaging. Furthermore, an optical gravitational arc, associated with the strongest lensed component, has been detected. Surrounding the system are a number of faint galaxies which may help explain the wide image separation. A plausible mass distribution model for CLASS B2108+213 is also presented.
\end{abstract}


\section{Introduction}
\label{intro}

The Cosmic Lens All-Sky Survey (CLASS; \citealt{myers02,browne02}) is the largest statistically complete gravitational lens survey ever undertaken. A complete sample of 11~685 flat-spectrum radio sources ($S_{\nu} \propto \nu^{\alpha}$ where $\alpha \geq -0.5$ between 1.4 and 4.85~GHz) was observed with the Very Large Array (VLA) in A configuration at 8.46~GHz between 1994 and 1999. The 0.2-arcsec angular resolution provided by the VLA allowed the straight forward identification of galaxy-scale gravitational lenses, which typically have image separations of $\sim 1$~arcsec. CLASS sources were identified as promising gravitational lens candidates if the VLA 8.46~GHz image contained multiple compact components with Gaussian FWHM $\leq 170$~mas, the image separations were between $0.3 \leq \Delta\theta \leq 15$~arcsec, the total integrated flux-density was $S_{8.46} \geq 20$~mJy and the component flux-density ratio was $\leq 10:1$. The detection of compact structure and extended emission in the multiple components at higher resolutions with MERLIN (Multi-Element Radio Linked Interferometer Network) and the VLBA (Very Long Baseline Array) confirmed the lensing hypothesis for 21 gravitational lenses during the first three phases of CLASS (see \citealt{browne02} and references therein for a discussion of all the gravitational lenses discovered by CLASS). 

In this paper, we present observations of the gravitational lens candidate CLASS B2108+213, which was discovered during the fourth and final phase of CLASS. In Section \ref{radio} we describe the VLA, MERLIN and VLBA radio observations which led to the discovery of the gravitational lens candidate CLASS B2108+213. Follow-up optical and infrared imaging of the system and the surrounding field, taken with the {\it Hubble Space Telescope}, is presented in Section \ref{photometry}. Using the available observational data, the lensing hypothesis, along with an analysis of the lensing potential, is discussed in Section \ref{disc}. Finally, in Section \ref{conc} we summarise our findings and outline the future work we intend to carry out on this intriguing new gravitational lens from CLASS. For all calculations we adopt an $\Omega_{M}=0.3$, $\Omega_{\Lambda}=0.7$ flat-universe, with a Hubble parameter of $H_{0} = 70$~km~s$^{-1}$ Mpc$^{-1}$.

\section{Radio observations}
\label{radio}

\subsection{VLA 8.46~GHz observation}

CLASS B2108+213 was observed with the VLA in A configuration at 8.46~GHz as part of the final phase of CLASS on 1999 August 17. The total on-source integration time was $\sim 30$ seconds. A 3.3 second correlator integration time with two 50~MHz IFs were used. 3C286 was chosen as the primary flux-density calibrator and phase-referencing was carried out with JVAS B2059+034 (Jodrell Bank VLA Astrometric Survey; \citealt{patnaik92,browne98,wilkinson98}). The data were edited and calibrated in the standard way using {\sc aips} ({\sc a}stronomical {\sc i}mage {\sc p}rocessing {\sc s}oftware) and imaged within the Caltech VLBI difference mapping package ({\sc difmap}; \citealt{shepherd97}) using the CLASS mapping script \citep{myers02}. The $uv$-data were naturally weighted and elliptical Gaussian model components were fitted to each image. The rms noise level was 180~$\mu$Jy~beam$^{-1}$ and the beam size was $0.26 \times 0.23$~arcsec$^{2}$.

\begin{figure}
\begin{center}
\setlength{\unitlength}{1cm}
\begin{picture}(6,7.6)
\put(-0.9,-0.0){\includegraphics{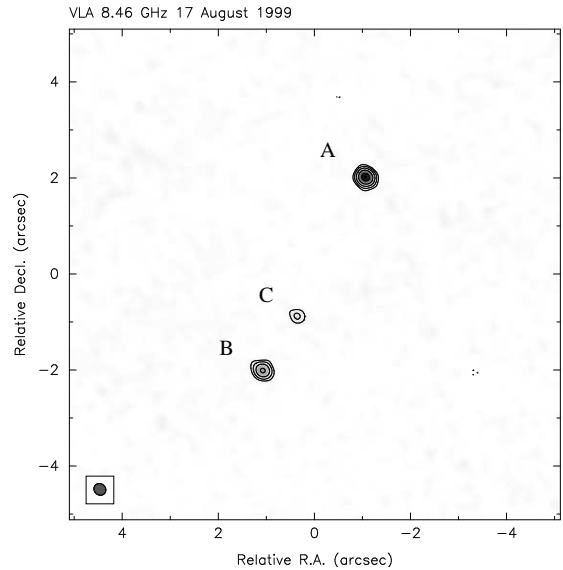}}
\end{picture}
\caption{VLA 8.46~GHz observation of CLASS B2108+213 taken during the final phase of CLASS on 1999 August 17. Three compact components were detected. The separation between components A and B is 4.56~arcsec. Contours are at $(-3,3,6,12,24,48) \times 180$~$\mu$Jy~beam$^{-1}$. The beam size is $259 \times 233$~mas$^2$ at a position angle of $50.6 \degr$. North is up and east is left.}
\label{vla-map}
\end{center}
\end{figure}

\begin{figure}
\begin{center}
\setlength{\unitlength}{1cm}
\begin{picture}(6,6.35)
\put(-2.0,6.8){\includegraphics{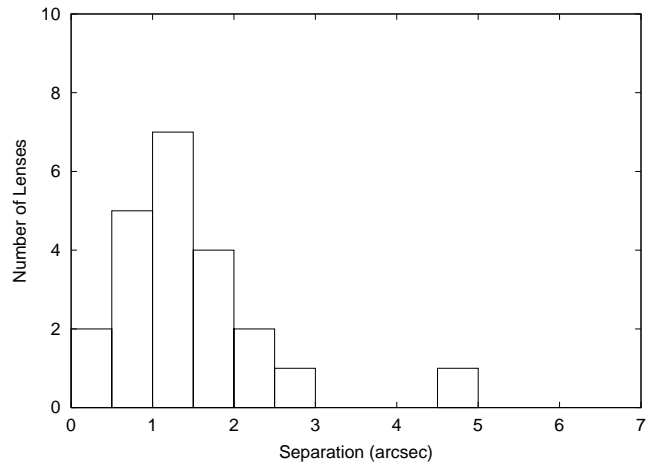}}
\end{picture}
\caption{The image separation distribution of the 22 CLASS gravitational lenses. CLASS B2108+213, with a maximum image separation of 4.56~arcsec, is the widest separation gravitational lens system in the CLASS sample.}
\label{image-sep}
\end{center}
\end{figure}

The VLA detected three compact components, shown in Fig. \ref{vla-map}, whose flux-densities and relative positions are given in Table \ref{data-tab} and Table \ref{2108-relative}, respectively. With an image separation between components A and B of 4.56~arcsec, CLASS B2108+213 has a wider image separation than any gravitational lens system thus far discovered by CLASS (see Fig. \ref{image-sep} for the CLASS gravitational lens image separation distribution). The total integrated flux-density of the system at 8.46~GHz is 18.5~mJy and the flux-density ratio between the brightest and faintest components is $\sim 8:1$. Therefore, CLASS B2108+213 failed to meet the strict lens candidate selection criteria due to its total integrated 8.46~GHz flux-density being less than 20~mJy. However, CLASS B2108+213 was followed up as a promising lens candidate even though it was no longer part of the statistically well defined CLASS sample. 

\begin{table*}
\begin{center}
\begin{tabular}{llrrrrrrc}
\hline
Date    & Instrument & Frequency & Integration & \multicolumn{1}{c}{Beam size}                             & \multicolumn{3}{c}{Flux density} & \multicolumn{1}{c}{$\sigma_{map}$}\\
        &            & \multicolumn{1}{c}{[GHz]}     & time [min]  & \multicolumn{1}{c}{[arcsec$^2$, P.A.]}  & \multicolumn{1}{c}{A [mJy]} & \multicolumn{1}{c}{B [mJy]}     & \multicolumn{1}{c}{C [mJy]} & \multicolumn{1}{c}{[$\mu$Jy~beam$^{-1}$]}    \\ \hline
1999 Aug 17 & VLA        &  8.460     & 0.5           & $0.259\times0.233$, $+51\degr$    & $11.7\pm0.6$ & $5.3\pm0.3$ & $1.5\pm0.2$ & 180\\
2000 Mar 10 & MERLIN     &  4.994     & 100.0         & $0.063\times0.049$, $+32\degr$    &  $8.8\pm0.5$ & $4.4\pm0.4$ & $<1.5$      & 298\\
2001 Mar 21 & VLBA       &  4.987     & 100.0         & $0.004\times0.002$, $-18\degr$    &  $6.8\pm0.4$ & $2.6\pm0.2$ & $<0.7$      & 130\\
2001 Aug 10 & MERLIN     &  4.994     & 540.0         & $0.070\times0.044$, $+25\degr$    & $10.6\pm1.6$ & $5.0\pm0.8$ & $1.1\pm0.2$ & 109\\
2001 Dec 31 & VLA        &  4.860     & 2.0           & $28.10\times15.00$, $-66\degr$    & $30.0\pm1.5$ & $\leftarrow$& $\leftarrow$& 173\\
2001 Dec 31 & VLA        &  8.460     & 2.0           & $14.90\times8.390$, $-60\degr$    & $25.8\pm1.3$ & $\leftarrow$& $\leftarrow$& 110\\
2001 Dec 31 & VLA        & 14.940     & 4.0           & $9.500\times4.810$, $-66\degr$    & $21.9\pm1.1$ & $\leftarrow$& $\leftarrow$& 303\\
2001 Dec 31 & VLA        & 22.460     & 7.0           & $5.950\times3.130$, $-61\degr$    & $10.6\pm0.6$ & $7.3\pm0.5$ & $\leftarrow$& 305\\
\hline
\end{tabular}
\caption{A summary of the radio observations of CLASS~B2108+213. Non-detections are quoted at the $5\sigma$ level. For those observations where the resolution was insufficient to separate components the flux-density is listed under the brightest component and the others are marked $\leftarrow$.}
\label{data-tab}
\end{center}
\end{table*}

\subsection{MERLIN 5~GHz observations}
\label{merlin-label}

In order to determine if any of the compact components detected by the VLA showed extended emission at higher resolutions we observed CLASS B2108+213 with MERLIN at 5~GHz for $\sim 1.5$~hours on 2000 March 10 and $\sim 9$~hours on 2001 August 10. The same set-up was used for both observations. 3C286 was used as the primary flux-density calibrator and JVAS B2103+213 was used for phase-referencing. The point source OQ208 was used to determine the antenna-based phase corrections. The observations were carried out using the left and right circular polarizations with $16 \times 1$~MHz channels. The data were initially edited and flux-density calibrated using the standard MERLIN data reduction programmes {\sc dplot} and {\sc tdproc} before being processed within {\sc aips} using the MERLIN automated pipeline \citep{thomasson94}. As before, the mapping was carried out within {\sc difmap}, using natural weighting. The rms noise level and beam size were 298~$\mu$Jy~beam$^{-1}$ and $63 \times 49$~mas$^2$ on 2000 March 10, and 109~$\mu$Jy~beam$^{-1}$ and $70 \times 44$~mas$^2$ on 2001 August 10. The observing conditions on 2001 August 10 were poor. As a consequence, an uncertainty of $\sim 15$~per~cent has been assigned to the flux-density calibration for that observation.

\begin{figure}
\begin{center}
\setlength{\unitlength}{1cm}
\begin{picture}(6,7.6)
\put(-0.9,-0.0){\includegraphics{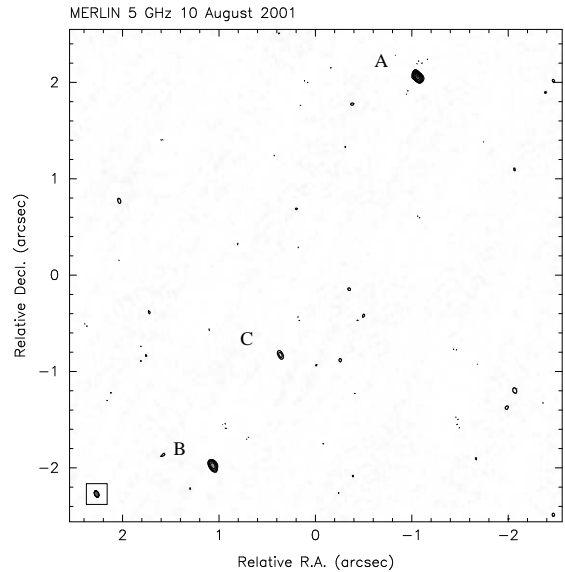}}
\end{picture}
\caption{MERLIN 5~GHz long-track observation of CLASS~B2108+213 taken on 2001 August 10. The FWHM of the elliptical Gaussian model fitted to component C is comparable to the MERLIN beam size. Contours are at $(-3, 3, 6, 12, 24, 48)  \times 109~\mu$Jy~beam$^{-1}$. The beam size is $70.1\times44.1$~mas$^2$ at a position angle of $24.6\degr$. North is up and east is left.}
\label{merlin-map}
\end{center}
\end{figure}

Components A and B were detected by the 2000 March 10 observation and were found to be compact (Gaussian FWHM of 34 and 30~mas, respectively). However, component C was not detected. This was due to the radio emission from component C being extended and resolved out by the MERLIN beam, or having a flux-density below $S_{4.994} < 1.5$~mJy, the $5\sigma$ image sensitivity. To investigate these possibilities the deeper 5~GHz observation on 2001 August 10 was undertaken. All three components of CLASS B2108+213 that were identified by the VLA were detected (see Fig. \ref{merlin-map}) and fitted with single elliptical Gaussian model components. As previously observed, components A and B were found to have compact structures (Gaussian FWHM of 27 and 21~mas, respectively) consistent with lensing. However, component C was resolved (Gaussian FWHM of $64$~mas). The flux-densities of each component from both observations are given in Table \ref{data-tab} and their relative positions, taken from the 2001 August 10 observation, are presented in Table \ref{2108-relative}.

\begin{table}
\begin{center}
\begin{tabular}{lllrr} \hline
Date        & Instrument  & Comp. & \multicolumn{1}{c}{$\Delta\alpha$} & \multicolumn{1}{c}{$\Delta\delta$}\\
            &             &       & \multicolumn{1}{c}{[mas]}	       & \multicolumn{1}{c}{[mas]}	   \\ \hline
1999 Aug 17 & VLA         & A     &                           $0\pm13$ &     $0\pm13$ \\
            &             & B     &                        $2139\pm15$ & $-4027\pm11$ \\
            &             & C     &                        $1425\pm34$ & $-2892\pm34$ \\
2001 Mar 21 & VLBA        & A     &                           $0\pm1~~$  &    $0\pm1~~$ \\
            &             & B     &                        $2131\pm1~~$ & $-4032\pm1~~$ \\
2001 Aug 10 & MERLIN      & A     &                           $0\pm11$ &     $0\pm11$ \\
            &             & B     &                        $2122\pm11$ & $-4039\pm11$ \\
            &             & C     &                        $1422\pm13$ & $-2888\pm13$ \\ \hline

\end{tabular}
\end{center}
\caption{The relative positions of each radio component of CLASS B2108+213 as measured with the VLA, VLBA and MERLIN. The uncertainties in the relative positions have been calculated from the uncertainty in the position of the Gaussian fitted to each component (i.e. beam size$/$SNR). The J2000 position of component A, as measured with the VLBA, is 21$^h$10$^m$54.0242$^s$ $+$21$\degr$31$\arcmin$0.690$\arcsec$. The flux-densities of each component can be found in Table \ref{data-tab}.}
\label{2108-relative}
\end{table}

\subsection{VLBA 5~GHz observation}

\begin{figure*}
\begin{center}
\setlength{\unitlength}{1cm}
\begin{picture}(12,7.8)
\put(-2,-0.0){\includegraphics{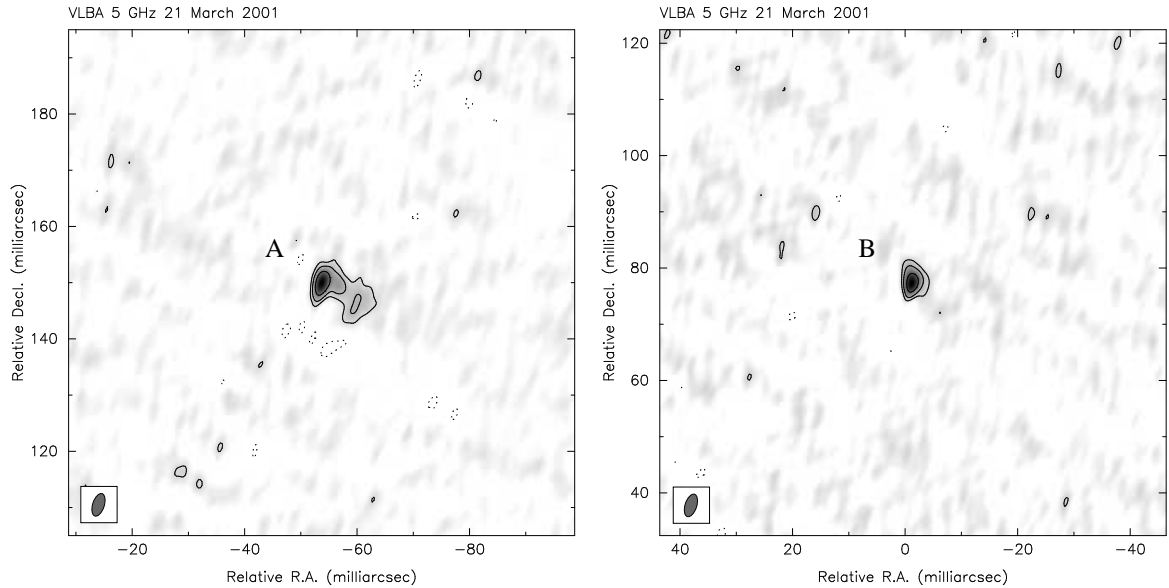}}
\end{picture}
\caption{Left: VLBA 5~GHz imaging of component A taken on 2001 March 21. Contours are at $(-3,3,6,12,24) \times 142$~$\mu$Jy~beam$^{-1}$. The beam size is $4\times2$~mas$^2$ at a position angle of $-18.1 \degr$. Right: The corresponding VLBA image of component B taken on 2001 March 21. Contours are at $(-3,3,6,12,24) \times 124$~$\mu$Jy~beam$^{-1}$. The beam size is $4 \times 2$~mas$^2$ at a position angle of $-18 \degr$. North is up and east is left.}
\label{vlba-map}
\end{center}
\end{figure*}

A snapshot observation with the VLBA at 5~GHz was made of CLASS B2108+213 on 2001 March 21. To provide uniform $uv$-coverage the observation was split into 5 separate 20 minute on-source integrations. Phase-referencing and fringe-fitting were carried out using JVAS B2103+213 and 3C345, respectively. The data were recorded in the left circular polarization through 4 IFs, each with 8~MHz bandwidth.\footnote{Note that each IF was split into $16\times0.5$~MHz channels during the data processing at the correlator.} Since the separation of components A and B was known to be 4.56~arcsec we obtained two correlations for the data with each centred on the respective components. The calibration and data editing was performed in the standard way using {\sc aips} before being averaged into 2~MHz frequency channels and 10 second integrations. The data were {\it clean}ed and mapped in {\sc aips} using {\it imagr} which allowed the data to be imaged without further frequency channel averaging. The rms noise level was 142 and 124~$\mu$Jy~beam$^{-1}$ for the respective maps of components A and B, which are given in Fig. \ref{vlba-map}. The beam size was $4 \times 2$~mas$^2$.

As before, both A and B components were detected. The brighter component (A) contains weak extended emission stretching to the south-west whereas component B is compact. Furthermore, the extension observed in component A is perpendicular to the axis connecting components A and B. This is consistent with the lensing hypothesis. However, component C was not detected. The sensitivity of these deep VLBA data should have been sufficient to detect component C if it were compact. The individual component flux-densities are given in Table \ref{data-tab} and the relative positions are given in Table \ref{2108-relative}.

\subsection{VLA multi-frequency observations}

Flux-density measurements of CLASS~B2108+213 at 4.86, 8.46, 14.94 and 22.46 GHz were carried out with the VLA in D configuration on 2001 December 31. CLASS~B2108+213 was observed for 2~minutes at 4.86 and 8.46~GHz, 4~minutes at 14.94~GHz and 7~minutes at 22.46~GHz. A 3.3~second correlator integration time was used with two 50~MHz IFs. 3C286 was used as the primary flux-density calibrator and phase-referencing was carried out using J2115+295, which was selected from the VLA calibrator list. The beam size and rms map noise at each observing frequency (4.86, 8.46, 14.94 and 22.46~GHz) were $28.1 \times 15.0$, $14.9 \times 8.4 $, $9.5 \times 4.8 $, $6.0 \times 3.1 $~arcsec$^2$ and 173, 110, 303, 305~$\mu$Jy~beam$^{-1}$, respectively.

The data were calibrated and edited in the standard way using {\sc aips}. Mapping and self-calibration were carried out within {\sc difmap}. Natural weighting was used to maximize the signal-to-noise of the maps. For the 4.86, 8.46 and 14.94~GHz observations CLASS B2108+213 was unresolved, therefore a single elliptical Gaussian model component was fitted to the data. However, the beam size of the 22.46~GHz observation was sufficient to resolve component A from components B and C. Therefore, two elliptical Gaussian model components were fitted to the positions of A and B/C. The angular resolution was not sufficient to resolve components B and C. At each observing frequency, strong emission was detected from CLASS~B2108+213 (see Table \ref{data-tab} for the respective flux-densities).

\section{Optical \& Infrared Observations}
\label{photometry}

An important step in the lens candidate confirmation process is the optical identification of a suitable lensing galaxy and the lensed images. This is especially important for the case of CLASS~B2108+213 since the wide image separation of 4.56~arcsec probably requires the presence of a group or cluster of lensing galaxies. 

High resolution imaging of CLASS~B2108+213 and the surrounding field was carried out with the {\it Hubble Space Telescope} (GO-9744) on 2003 September 09. The optical data were taken with the Advanced Camera for Surveys (ACS), in the Wide Field Channel (WFC) mode, which has a field of view of $202 \times 202$ arcsec$^2$ and a pixel scale of 0.05~arcsec pixel$^{-1}$. The total on-source integrations, taken through the F555W and F814W filters, were 2.2~ks and 2.3~ks, respectively. Infrared imaging through F160W was carried out with the Near Infrared Camera/Multi-Object Spectrograph (NICMOS). This 2.6~ks observation was taken with the NIC2 camera in the multi-accum data taking mode. NIC2 has a field of view of $19.2 \times 19.2$ arcsec$^2$ and a pixel scale of 0.075~arcsec pixel$^{-1}$.

\begin{figure*}
\begin{center}
\setlength{\unitlength}{1cm}
\begin{picture}(12,11.8)
\put(-2.8,-0.1){\includegraphics{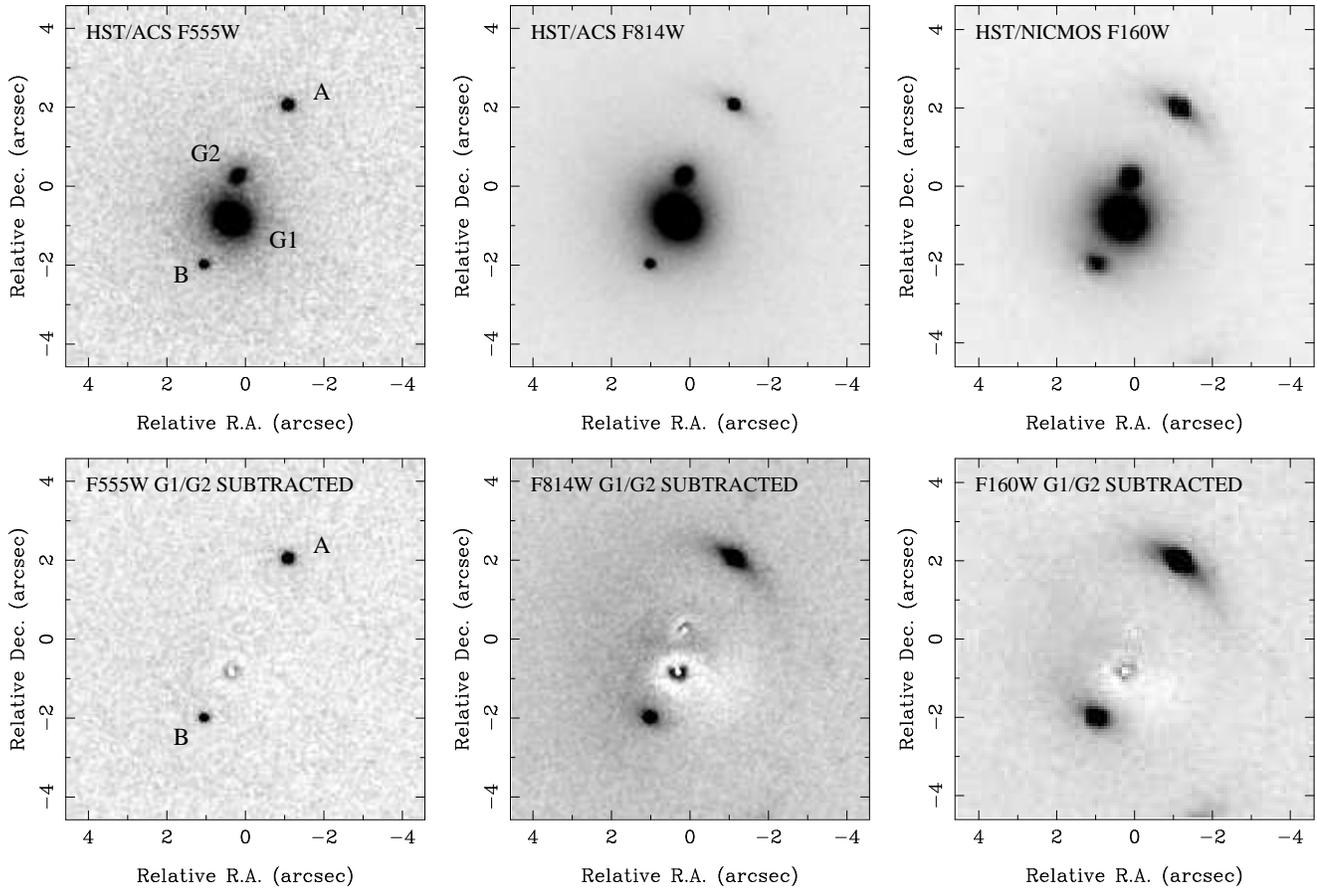}}
\end{picture}
\caption{Optical and infrared imaging of CLASS B2108+213 taken with the {\it Hubble Space Telescope} on 2003 September 09. Both lensed images (A and B) and two lensing galaxies (G1 and G2) have been detected. Top, from left to right, are the ACS F555W, ACS F814W and NICMOS F160W images, respectively. The same images, with the lensing galaxies subtracted, are shown on the bottom. The fitted galaxy profiles are a good approximation of the core emission from both lensing galaxies (i.e. observed/model relative residuals of $\sim$~4 percent). In the ACS F555W image the lensed components appear point-like. Although a strong point source was detected, a faint underlying arc feature, which is presumably from the host galaxy, can also be seen in the ACS F814W and NICMOS F160W images of component A. North is up and east is left.}
\label{hst}
\end{center}
\end{figure*}

The data were reduced using standard {\it stsdas} tasks within {\sc iraf}.\footnote{{\sc iraf} ({\sc i}mage {\sc r}eduction and {\sc a}nalysis {\sc f}acility) is distributed by the National Optical Astronomy Observatories, which are operated by AURA, Inc., under cooperative agreement with the National Science Foundation.} The ACS data for each filter were first processed using {\it calacs}, before being combined and corrected for the ACS geometric distortion using {\it multidrizzle} \citep{koekemoer02}. The infrared NICMOS data were reduced using the {\it calnica} and {\it calnicb} tasks. Further image processing (i.e. bad pixel and residual cosmic ray removal) was carried out using the {\it fillbad} and {\it clean} tasks within {\sc kappa} and {\sc figaro},\footnote{{\sc kappa}, {\sc figaro} and {\sc ccdpack} and are part of the {\sc starlink} project - http://www.starlink.rl.ac.uk} respectively. A median weighted mosiac of the background-subtracted dithered images was created using {\sc ccdpack}. The ACS F555W and F814W, and NICMOS F160W images of CLASS B2108+213 are presented in Fig. \ref{hst} (top).

\begin{figure}
\begin{center}
\setlength{\unitlength}{1cm}
\begin{picture}(6,8.0)
\put(-1.30,-0.40){\includegraphics{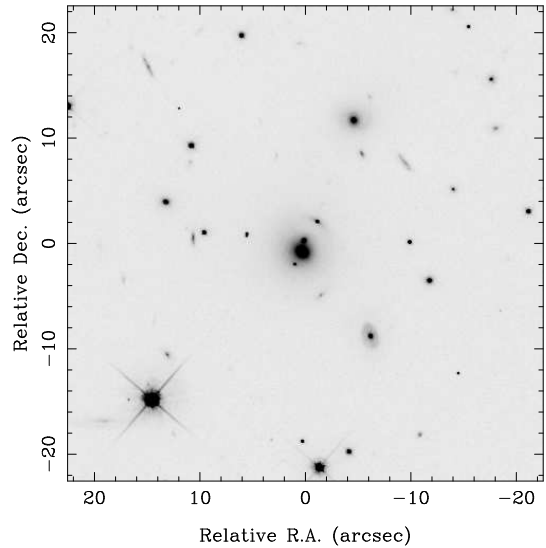}}
\end{picture}
\caption{{\it Hubble Space Telescope} ACS F814W imaging of the CLASS B2108+213 field. Surrounding the lens system are a scatter of faint galaxies which could be contributing to the lensing potential. North is up and east is left.}
\label{hst-field}
\end{center}
\end{figure}

Optical and infrared emission coincident with the two lensed images (A and B) has been detected. Both lensed images appear to be strong point sources, however, extended arc-like emission, which is presumably from the host galaxy, was also detected in the F814W and F160W images of component A. A lensing galaxy (G1) has been found within 6~mas of the position of radio component C (based on the MERLIN 5~GHz position). Faint emission from a fourth component (G2) within the Einstein radius of the system is likely to be from a companion lensing galaxy. Both deflectors have a morphology which is consistent with an early-type galaxy, although there is evidence that G1 may be disturbed (see below for surface brightness profile fitting). In Fig. \ref{hst-field} we show the ACS F814W image of the CLASS B2108+213 field. Surrounding the lens system are a scatter of faint galaxies which may be contributing to the gravitational potential and might help account for the wide image separation. However, we defer any discussion of a possible group associated with CLASS B2108+213 to a follow-up paper (McKean et al., in preparation).

In order to carry out a photometric and astrometric analysis of the data, the surface brightness profiles of the two lensing galaxies were fitted with a de Vaucouleurs' profile, convolved with the ACS/NICMOS point spread function,\footnote{The ACS and NICMOS point spread functions were generated using {\sc tinytim} \citep{krist95}.} using {\sc galfit} \citep{peng02}. The parameters of the profiles fitted to the ACS F814W data-set, which has the highest signal-to-noise ratio of the three images, are given in Table \ref{galfit} and the residual images, with the lensing galaxies subtracted, are shown in Fig. \ref{hst} (bottom). The fitted profiles are a good representation of the galaxy cores, with relative residuals (i.e. observed/model) of $\sim$~4 percent. However, one of the most striking features of the lensing galaxy subtracted images is the asymmetry in the residuals in the outer tail of G1. The fitted surface brightness profile over (under) represents the observed profile toward the west (east). The likely cause of these asymmetric residuals is the interaction between the lensing galaxies which could lead to the disturbed morphology of G1.

The magnitudes of the lensed components were measured within elliptical apertures using the images with the lensing galaxies subtracted. The same size of aperture was used to calculate the magnitudes of A and B in the F555W, F814W and F160W images. The positions of the lensed components were found by fitting point sources to the strong core emission detected in each image. The pixel scales for each instrument, which are required for the astrometric analysis of the data, were established by assuming that the radio and optical core emission from components A and B was coincident. Using the precise AB image separation provided by the VLBA 5~GHz observations, we found the (x, y) pixel scales (mas pixel$^{-1}$) for the ACS and NICMOS to be (50.320, 49.983) and (75.929, 75.000), respectively. The relative positions and magnitudes of each component are presented in Table \ref{hst-flux}. Note that the magnitudes quoted in Table \ref{hst-flux} have not been corrected for galactic extinction.

\begin{table}
\begin{center}
\begin{tabular}{llrrc} \hline
Filter      & Comp.     & \multicolumn{1}{c}{$\Delta\alpha$}  & \multicolumn{1}{c}{$\Delta\delta$}   & Magnitude     \\ 
            &           & \multicolumn{1}{c}{[mas]}        & \multicolumn{1}{c}{[mas]}         & [Vega]        \\ \hline
F555W       & A         & $   0\pm1$ &     $0\pm1$ & $22.56\pm0.15$\\
            & B         & $2131\pm3$ & $-4032\pm4$ & $23.73\pm0.17$\\
            & G1        & $1427\pm5$ & $-2890\pm5$ & $19.91\pm0.11$\\
            & G2        & $1273\pm5$ & $-1786\pm5$ & $22.59\pm0.15$\\
F814W       & A         & $   0\pm1$ &     $0\pm1$ & $20.96\pm0.15$\\
            & B         & $2131\pm2$ & $-4032\pm3$ & $21.97\pm0.17$\\
            & G1        & $1428\pm2$ & $-2888\pm2$ & $17.63\pm0.11$\\
            & G2        & $1275\pm2$ & $-1786\pm2$ & $20.61\pm0.15$\\
F160W       & A         & $   0\pm1$ & $    0\pm1$ & $18.14\pm0.15$\\
            & B         & $2131\pm2$ & $-4032\pm3$ & $18.86\pm0.17$\\
            & G1        & $1421\pm2$ & $-2888\pm2$ & $15.65\pm0.11$\\
            & G2        & $1263\pm1$ & $-1795\pm2$ & $18.36\pm0.15$\\ \hline
\end{tabular}
\end{center}
\caption{The magnitudes (Vega) and relative positions of each optical component of CLASS B2108+213.}
\label{hst-flux}
\end{table}

\begin{table}
\begin{center}
\begin{tabular}{lll} \hline
Parameter        & G1                       & G2          \\ \hline
Total magnitude  & 17.63                    & 20.61       \\
Effective radius & 1.36~arcsec              & 0.19~arcsec \\
Axial ratio      & 0.86                     & 0.79        \\
Position angle   & 57$\degr$                & 134$\degr$  \\ \hline
\end{tabular}
\end{center}
\caption{The model parameters of the de Vaucouleurs' profiles fitted to the ACS F814W data of G1 and G2 using {\sc galfit} \citep{peng02}. These parameters are consistent with those fitted to the ACS F555W and NICMOS F160W data. The position angles are measured east of north.}
\label{galfit}
\end{table}

\section{Discussion}
\label{disc}

The observational evidence presented above will now be used to argue for the gravitational lensing hypothesis in CLASS B2108+213 and investigate the nature of the third radio component. A plausible model for the lensing potential is also presented.

\subsection{Evidence for the lensing hypothesis}

The radio data presented in Section \ref{radio} provide compelling evidence for the lensing hypothesis in CLASS B2108+213. Firstly, the VLA and MERLIN images have identified two compact components (A and B) separated by 4.56~arcsec with flat radio spectra, $\alpha_{5}^{8.46} \sim 0.15$. High resolution imaging with the VLBA has shown that component A exhibits extension stretching toward the south-west, which is perpendicular to the axis connecting components A and B; whereas the weaker component, B, is compact. This is typical of the gravitational lensing phenomena as surface brightness in the lensed images is conserved (e.g. \citealt*{schneider92}). Also, as gravitational lensing is achromatic and radio emission is not affected by dust absorption, we would expect the flux-density ratios of multiple images of the same background source to be nearly identical over all frequencies. The flux-density ratios of components A and B at 5 and 8.46~GHz are $S_{5,B}/S_{5,A}=0.47 \pm 0.10$ and $S_{8.46,B}/S_{8.46,A}=0.45 \pm 0.03$ respectively. Furthermore, from the {\it Hubble Space Telescope} NICMOS F160W data presented in Section \ref{photometry}, the infrared flux-density ratio\footnote{This flux-density ratio has been calculated using only the emission from the strong point sources.} of components A and B was found to be $S_{F160W,B}/S_{F160W,A} = 0.51\pm0.04$. Such similar flux-density ratios between 6~cm and 1.6~$\mu$m provides strong confirmatory evidence that components A and B are lensed images. Further evidence for the lensing hypothesis is provided by the under-lying gravitational arc detected in the F814W and F160W images of component A. The identification of two suitably positioned lensing galaxies also adds to the case for components A and B being images of the same background source.

\subsection{The nature of the third radio component}

One of the most interesting features of CLASS B2108+213 is the possible detection of a third lensed image (although, the observations presented here suggest the emission from component C is probably from the lensing galaxy, G1). Lensing theory predicts that for extended mass distributions there should exist an odd (i.e. 3rd, 5th, etc) lensed image which is positioned very close to the centre of the lensing potential \citep{dyer80,burke81}. Therefore, the detection (or non-detection) of such an image can be used to constrain the core-properties of early-type galaxies (\citealt*{rusin01}; \citealt{keeton03}). Unfortunately, being positioned near to the lens centre means that odd images are also highly de-magnified. As a consequence their detection is extremely rare. The triple component gravitational lens system PMN~J1632-0033 (\citealt{winn02}; \citealt*{winn03,winn03a}) currently has the best core image candidate. Other gravitational lenses which were found to have a candidate odd-image have subsequently shown that the emission is probably not from a lensed image but from a lensing galaxy (e.g. \citealt{chen93,fassnacht99a}).

\begin{figure}
\begin{center}
\setlength{\unitlength}{1cm}
\begin{picture}(6,6.55)
\put(-0.7,-0.3){\includegraphics{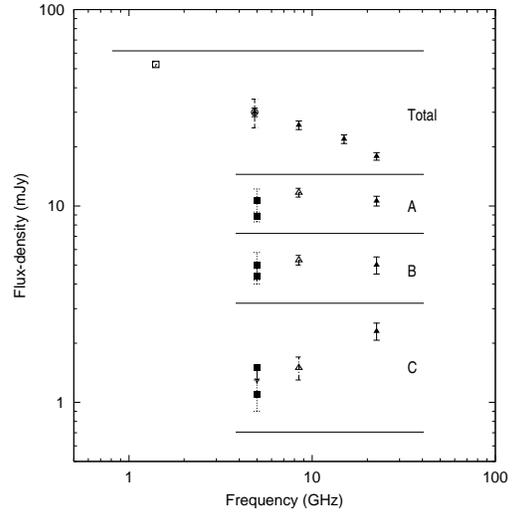}}
\end{picture}
\end{center}
\caption{The radio spectral energy distribution of CLASS B2108+213 and components A, B and C. The empty square and circle are the NVSS \citep{condon98} and GB6 \citep{gregory96} flux-densities of CLASS B2108+213 respectively. The empty and filled triangles represent the CLASS VLA and multi-frequency VLA flux-densities. The flux-densities measured by MERLIN are the filled squares. Notice that the GB6 and VLA $\sim 4.85$~GHz flux-densities are identical. The 22.46~GHz flux-densities for components B and C have been estimated by assuming that the flux-density ratio, $S_{B}/S_{A}$ is constant. Note that the sum of the radio component flux-densities is less than the total flux-density measured by GB6 and the VLA multi-frequency observations. This is most apparent at $\sim 4.85$~GHz and is probably due to resolution effects.}
\label{spec}
\end{figure}

The latter is almost certainly the case in CLASS B2108+213 for two reasons. Firstly, from the long-track MERLIN 5~GHz image presented in Section \ref{merlin-label}, it is apparent that component C is extended (recall that the FWHM of the elliptical Gaussian fitted to C is comparable to the MERLIN beam size) whereas components A and B are both compact. This conclusion is consistent with the non-detection of component C with the VLBA at 5~GHz. Secondly, the implied radio spectrum of component C differs from that of A and B. As can be seen in Fig. \ref{spec}, component C has an inverted radio spectrum between 5 and 22.46~GHz, whereas A and B show a turnover.\footnote{The flux-density of component C at 22.46~GHz has been estimated by assuming that the flux-density ratio of A and B at 22.46~GHz is the same as at 5 and 8.46~GHz.} Therefore, the component C radio emission is probably from the lensing galaxy, G1, because the emission is not compact and has a different radio spectrum from the lensed images, A and B. There is a subtle flaw to this argument. The implied radio spectrum of component C is rising at 22.46~GHz. This suggests that the source of the component C radio emission is very compact. However, the MERLIN 5~GHz imaging has shown component C to be extended. A possible explanation for this contradiction is scattering in the lensing galaxy, which could decrease the surface brightness of the radio emission (e.g. \citealt{guirado99,biggs03a,winn03b,biggs04}). However, accepting a scatter broadening explanation for the observed extension opens up the possibility that component C may be a lensed image since surface brightness will no longer be conserved.

Further radio observations to determine the actual radio spectra of the CLASS B2108+213 radio components will need to be carried out to resolve this issue. However, the most likely scenario is that the radio emission from component C is from the lensing galaxy. Although this conclusion rules out the exciting possibility of using CLASS B2108+213 for studying the core properties of the lensing potential, an accurate position for the lensing galaxy G1 has probably been obtained. This will improve the mass model of CLASS B2108+213 which is discussed in the following section.

\subsection{Modelling the lensing potential}

Using the observational data presented in this paper, a preliminary analysis of the CLASS B2108+213 lensing potential can now be carried out. The lensed image positions and flux-densities provided by the VLBA 5~GHz data (see Table \ref{data-tab} and Table \ref{2108-relative}) give five model constraints; two from each image position and one from the flux-density ratio. The positions of the lensing galaxies have been taken from the NICMOS F160W data and allowed to vary within the quoted uncertainties (see Table \ref{hst-flux}). Each lensing galaxy is represented by a singular isothermal sphere, which has a convergence ($\kappa$) of,
\begin{equation}
\kappa = \frac{b}{2r},
\end{equation}
where $b$ is the Einstein radius and $r$ is the radial distance from the lens centre. For a singular isothermal sphere the Einstein radius is given by,
\begin{equation}
b = 4\pi \left( \frac{\sigma_{v}}{c} \right)^2 \frac{D_{ls}}{D_{s}},
\end{equation}
where $\sigma_{v}$ is the one-dimensional velocity dispersion, and $D_{ls}$ and $D_{s}$ are the angular diameter distances to the background source from the lens and observer, respectively. The ratio of the G1 and G2 luminosities ($L$) has been used to constrain the mass ratio of the lensing galaxies ($b_{G2}/b_{G1}=0.29\pm0.05$) by invoking the Faber-Jackson relation ($L \propto b^2$ for a singular isothermal sphere; \citealt*{faber76}). A mass model containing two singular isothermal spheres, with the mass ratio fixed, requires three free parameters; two for the source position and one for the Einstein radius of G1. The final two free parameters have been used for the position angle and strength of an external shear.

\begin{figure}
\begin{center}
\setlength{\unitlength}{1cm}
\begin{picture}(6,6.97)
\put(-1.0,-0.3){\includegraphics{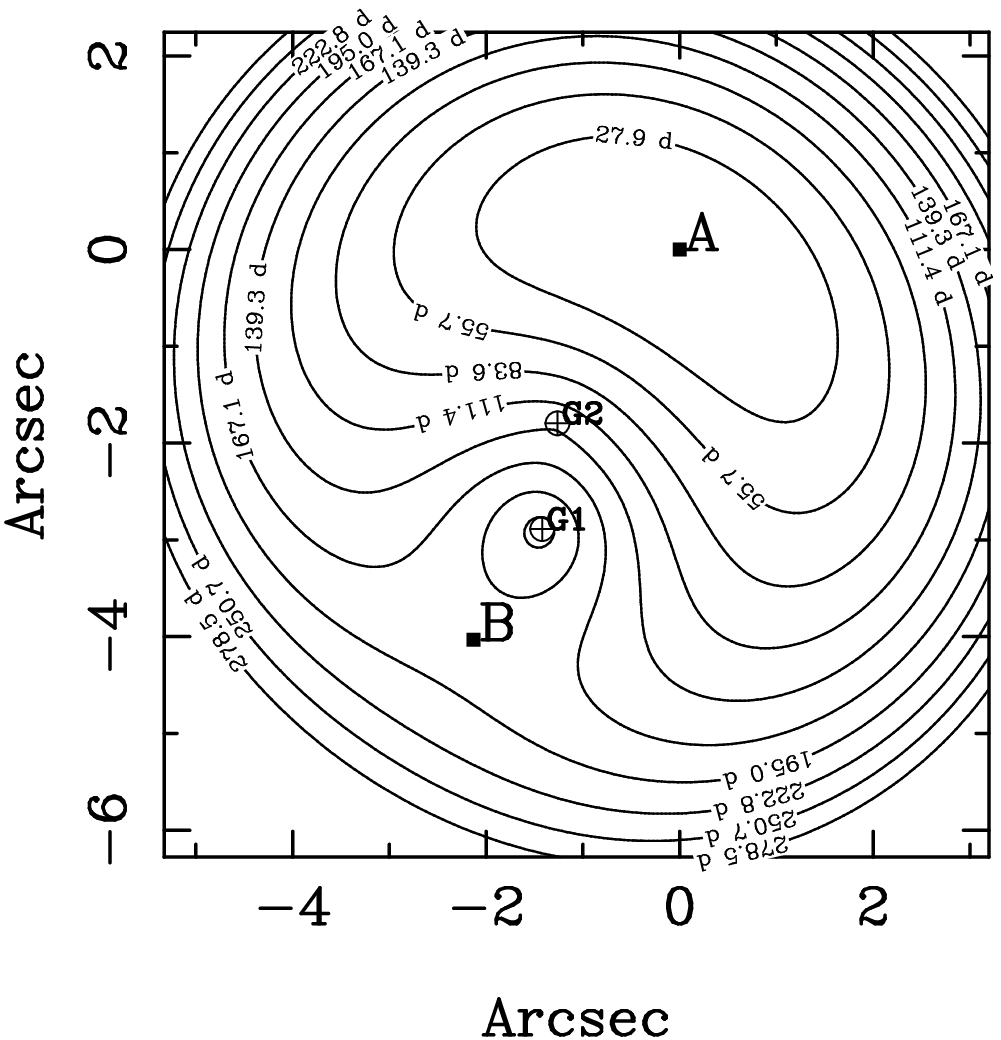}}
\end{picture}
\end{center}
\caption{The time-delay surface for the CLASS B2108+213 mass model. The model consists of two singular isothermal spheres and an external shear. The time-delay is measured in days and assumes that G2 is at the same redshift as G1 ($z_{G1}=0.365$; McKean et al., in preparation) and that the source redshift is 1.5. However, since the source redshift is unknown, the cosmology dependent time-delay is only an estimate. For different source redshifts (or Hubble constant), each contour level scales by a constant factor. The model parameters are given in Table \ref{model-tab}.}
\label{model}
\end{figure}

The time-delay surface of our mass model for CLASS B2108+213 is shown in Fig. \ref{model} and the model parameters are given in Table \ref{model-tab}. We find that two singular isothermal spheres, with an external shear component, reproduces the observed positions and flux-densities of the lensed images. 

Due to the limited number of observational constraints, this mass model has no degrees of freedom. Further improvements of this simple model, which only serves to show that the system can be explained by gravitational lensing, will be presented in a companion paper (McKean et al., in preparation). Additional constraints from stellar kinematic data for G1, the extended structure of the lensed images, and redshift and luminosity data on putative group members around the lens system will be used to further investigate the CLASS B2108+213 lensing potential.

\begin{table}
\begin{center}
\begin{tabular}{lll} \hline
Comp.  & \multicolumn{2}{c}{Parameter} \\ \hline
G1     & $\alpha,~\delta$     & 1421, $-$2888~mas\\
       & $b_{G1}$             & 1786~mas\\
G2     & $\alpha,~\delta$     & 1263, $-$1795~mas\\
       & $b_{G2}$             & 518~mas\\
Shear  & Magnitude            & 0.018\\
       & Position angle       & 118.1$\degr$\\
A      & $\alpha,~\delta$     & 0, 0~mas\\ 
       & Flux density (5~GHz) & 6.7~mJy\\
B      & $\alpha,~\delta$     & 2131, $-$4032~mas\\ 
       & Flux density (5~GHz) & 2.6~mJy\\
Source & $\alpha,~\delta$     & 1028, $-$2033~mas\\ 
       & Flux density (5~GHz) & 1.4~mJy\\ \hline
\end{tabular}
\end{center}
\caption{The parameters of the CLASS B2108+213 mass model. The model successfully reproduces the observed positions and flux-densities of the lensed components (see Tables \ref{data-tab} and \ref{2108-relative}). All positions are measured relative to component A.}
\label{model-tab}
\end{table}

\section{Conclusions and Further Work}
\label{conc}

We have presented radio and optical observations of the new gravitational lens CLASS B2108+213. This system comprises two lensed images, which are separated by 4.56~arcsec, thus making CLASS B2108+213 the widest separation gravitational lens discovered by CLASS. Optical imaging found two lensing galaxies within the Einstein radius of the system. Intriguingly, we also detected radio emission from a third component which was coincident with the optical emission from the primary lensing galaxy. The extension of the third radio component and its different radio spectrum from the two lensed images suggests that it is unlikely to be a third lensed image. However, it is too early to completely discount the possibility that scatter broadening effects might have been responsible for modifying the properties of a third image. The positions and flux-densities of the lensed images were successfully reproduced by a model containing two singular isothermal spheres and an external shear. However, further observational constraints will need to be obtained for a detailed analysis of the CLASS B2108+213 lensing potential to be carried out.

The wide image separation of CLASS B2108+213 implies a group/cluster assisted lensing potential. Therefore, CLASS B2108+213 provides an excellent opportunity to study the structure and mass distribution of galaxy groups beyond the local universe ($z > 0.2$). We have begun a spectroscopic programme to determine the redshifts of the lensing galaxies and lensed source in CLASS B2108+213, and to identify which nearby galaxies are part of the lensing group. Thus far we have found four nearby galaxies at similar redshifts to G1 ($z_{G1} = 0.365$; McKean et al., in preparation). Furthermore, our spectrum of G1 has sufficient signal-to-noise for the stellar velocity dispersion to be measured. These observations will provide additional constraints for a future mass model.

The other interesting feature of CLASS B2108+213 is the possible detection of a third lensed image. Although the observations presented here show that this is almost certainly not the case, this conclusion could be confirmed by measuring the flux-density, polarization and size of the radio emission from each component as a function of frequency. If scattering of a third lensed image is occurring then we would expect the high frequency spectrum of each radio component to be identical.

\subsection*{ACKNOWLEDGMENTS}
The VLA and VLBA are operated by the National Radio Astronomy Observatory which is a facility of the National Science Foundation operated under cooperative agreement by Associated Universities, Inc. MERLIN is a national facility operated by the University of Manchester on behalf of PPARC. The results present herein were based on observations collected with the NASA/ESA {\it HST}, obtained at STScI, which is operated by AURA, under NASA contract NAS5-26555. JPM, MAN, PMP and TDY acknowledge the receipt of PPARC studentships. RDB is supported by NSF grant AST-9900866. TT acknowledges support from NASA through Hubble Fellowship grant HST-HF-01167.01-A. This work was supported by the European Community's Sixth Framework Marie Curie Research Training Network Programme, Contract No. MRTN-CT-2004-505183 ``ANGLES''.

\bsp

\label{lastpage}

\end{document}